# Thermal Phase Shifters for Femtosecond Laser Written Photonic Integrated Circuits

Francesco Ceccarelli, Simone Atzeni, Alessandro Prencipe, Raffaele Farinaro, and Roberto Osellame, *Fellow, OSA*

*Abstract*—Photonic integrated circuits (PICs) are today acknowledged as an effective solution to fulfill the demanding requirements of many practical applications in both classical and quantum optics. Phase shifters integrated in the photonic circuit offer the possibility to dynamically reconfigure its properties in order to fine tune its operation or to produce adaptive circuits, thus greatly extending the quality and the applicability of these devices. In this paper, we provide a thorough discussion of the main problems that one can encounter when using thermal shifters to reconfigure photonic circuits. We then show how all these issues can be solved by a careful design of the thermal shifters and by choosing the most appropriate way to drive them. Such performance improvement is demonstrated by manufacturing thermal phase shifters in femtosecond laser written PICs (FLW-PICs), and by characterizing their operation in detail. The unprecedented results in terms of power dissipation, miniaturization and stability, enable the scalable implementation of reconfigurable FLW-PICs that can be easily calibrated and exploited in the applications.

## I. INTRODUCTION

RECENT years have witnessed the rising of integrated photonics as an enabling tool in many emerging applications in both classical [1] and quantum optics [2]. In addition to the well-known advantages in terms of scalability, miniaturization and stability, a very important feature that photonic integrated circuits (PICs) provide to the applications is the possibility to reconfigure their behavior dynamically [3]–[6]. In particular, one of the basic building blocks of a reconfigurable PIC is the phase shifter [7], that is at the basis of many reconfigurable devices like tunable micro-ring resonators [8], phase-controlled Mach-Zehnder interferometers (MZIs) [9], optical switches [10] and many others. The advantage that reconfigurability can bring is not only the possibility to perform more than one experiment with the same photonic device: indeed, phase shifting is today exploited also for the compensation of the imperfections due to the PIC fabrication tolerances [3], for the development of 2D optical beam steering systems [4], for adaptive quantum sensing [5] and for the implementation of quantum machine learning protocols [6].

Phase shifting is based on the local variation of the refractive index of a waveguide. This effect can be achieved through different physical mechanisms, depending also on the integrated photonic platform adopted for the PIC fabrication. As an example, when using silicon photonics technology, phase shifters can be implemented by plasma dispersion effect, namely by modulating the concentration of free carriers within the waveguide core [11]. This approach allows one to reach modulation speeds in the gigahertz range, thus meeting the requirements of modern telecommunications. However, in many applications such a high bandwidth is not necessary. As an example, in quantum photonics experiments the measurement time is usually dominated by the limited generation rate of the available multi-photon sources [2]. Furthermore, free carriers injection leads to additional losses, which represent one of the bottlenecks of quantum experiments. Therefore, it is often more convenient to exploit the thermo-optic effect. By modulating the temperature of the waveguide with a resistive microheater, it is possible to induce a local modification of the refractive index and, in turn, a phase shift on the guided photons. Such a component is usually referred to as thermal shifter [12].

Thermal shifters have been demonstrated for several integrated photonics technologies, like silicon photonics [6], [13], silica-on-silicon [7], [14] and, more recently, also for platforms based on direct femtosecond laser writing (FLW) of waveguides in transparent substrates [12], [15]. In particular, the FLW technology has recently attracted a lot of attention thanks to its unique advantages [16]: firstly, since the material modification is confined to the focal volume, it is possible to realize complex 3D optical circuits by simply focusing the laser inside the sample and translating the latter with respect to the beam; secondly, this technique does not require a mask and, thus, it allows a rapid and cost-effective prototyping of circuits with different geometries.

However, the fabrication of a reconfigurable FLW-PIC featuring efficient thermal phase shifters is not a straightforward task. Indeed, in order to guarantee a proper thermal coupling between the microheaters and the circuit, it is necessary to write the waveguides as close as possible to the surface. On the other hand, when the laser is focused close to the surface, the process becomes very sensitive to the roughness and irregularities of the chip, resulting in defects and interruptions along the path of the fabricated waveguides. This issue limits the depth at which it is possible to write the circuit to few tens of micrometers. Moreover, silicate glasses, that are often considered the

Manuscript received April 3, 2019; revised June 3, 2019; accepted June 12, 2019. Date of publication June 14, 2019; date of current version August 16, 2019. This work was supported by the European Research Council under the European Union's Horizon 2020 research and innovation programme (project CAPABLE) under Grant 742745. *(Corresponding author: Francesco Ceccarelli.)*

The authors are with the Istituto di Fotonica e Nanotecnologie-Consiglio Nazionale delle Ricerche (IFN-CNR), 00185 Rome, Italy, and also with the Dipartimento di Fisica-Politecnico di Milano, 20133 Milano, Italy (e-mail: francesco.ceccarelli@polimi.it; simone.atzeni@polimi.it; alessandro.prencipe@mail.polimi.it; raffaele.farinaro@mail.polimi.it; roberto.osellame@polimi.it).

Color versions of one or more of the figures in this paper are available online at http://ieeexplore.ieee.org.

Digital Object Identifier 10.1109/JLT.2019.2923126

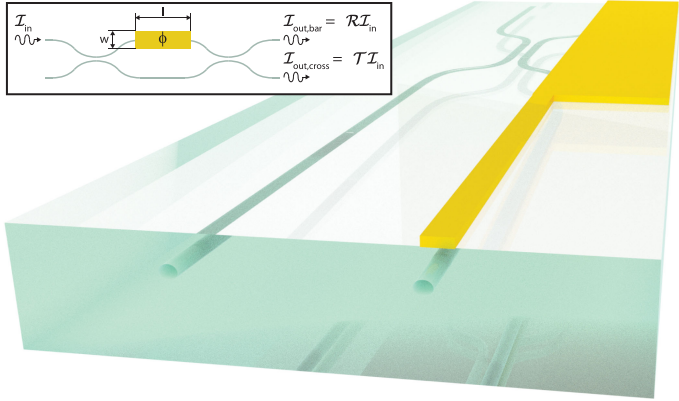

Fig. 1. 3D section of a reconfigurable MZI, depicting the device cut in half. Inset: 2D representation of the whole interferometer, in which the thermal shifter is reported along with its dimensions $l$ and $w$.

substrate of choice for the FLW-PIC fabrication, are characterized by a thermo-optic coefficient [17] that is more than one order of magnitude lower than other materials, typically employed for the fabrication of PICs, like silicon [13]. In this scenario, the electrical power needed for a complete phase shift of the optical signal can not be lower than many hundreds of milliwatts and, even micromachining the substrate with isolation trenches, this value has never been reduced to less than 200 mW [18]. This is a critical issue that can strongly affect the stability and even the reliability of the device. Moreover, it is worth noting that, in order to prevent the temperature drift of the whole chip and, in turn, to guarantee a reliable coupling with the external optical fibers and a stable value for the microheater resistivity, the total dissipated power must be limited to few watts. This issue prevents the integration of more than a few microheaters in the same chip without employing an active cooling system.

In this paper, we present a simple and effective strategy for the implementation of efficient thermal phase shifters in FLW-PICs. Thanks to the new technology, we demonstrate the lowest power dissipation reported in the literature for a MZI integrated in a glass substrate, with no need of more complex structuring as isolation trenches [18]. After that, we show that the current control of the thermal shifters has a beneficial effect both in terms of crosstalk and linearity, thus allowing an easier exploitation of FLW-PICs in the applications.

## II. Design Guidelines

Without loss of generality, let us consider an integrated planar MZI fabricated in a silicate glass and featuring a reconfigurable transmission coefficient $\mathcal{T}$ (Fig. 1). This means that, given an input signal characterized by an optical intensity $\mathcal{I}_{in}$ and assuming negligible losses, on the cross output the signal will be

$$\mathcal{I}_{out,cross} = \mathcal{T}\mathcal{I}_{in}, \quad (1)$$

while on the bar output we will expect an intensity

$$\mathcal{I}_{out,bar} = \mathcal{R}\mathcal{I}_{in}, \quad (2)$$

where $\mathcal{R} = 1 - \mathcal{T}$ is the reflection coefficient of the MZI. Reconfigurability is attained by placing a thermal shifter right above one of the two arms of the interferometer. The microheater can be realized with the process reported in [12]: firstly, a thin film of metal is deposited on the chip surface and, secondly, the resistors, the metal interconnections and the contact pads are laid out by ablating the film with the same femtosecond laser used to write the waveguides.

This process can be optimized by adopting the following design guidelines.

### A. Metal Film Design

A convenient choice for the microheater fabrication is the use of a gold film. Indeed, gold is not prone to the formation of native oxide and, therefore, the connection of the chip pads to an external power supply system (e.g., by wire-bonding) is straightforward. However, the adhesion of gold to a glass substrate is particularly weak and, thus, before depositing the gold layer, it is necessary to use a thin chromium film as an adhesion layer for the gold one [19].

The choice of the thickness for both the materials represents a critical step. Speaking of chromium, this layer can not be too thin, otherwise the adhesion to the substrate would be too weak; on the other hand, it cannot be too thick, otherwise, even considering an operating temperature up to 200 °C, the chromium atoms would start diffusing inside the gold layer [20], resulting in a drift of the film resistivity and, in turn, in an instability of the power injected during the heating. However, the latter issue can be solved by reducing the chromium thickness down to the minimum level that guarantees a good adhesion of gold to the substrate [19]. We verified with manual tape pull tests that this value is about 2 nm.

Conversely, given the substantial difference in resistivity between the two materials, the gold film is the layer that sets all the electrical properties of the microheaters. Therefore, the choice of the gold thickness $h$ must start from the relation

$$R_{ts} = \frac{\rho}{h}\frac{l}{w} = R_{sh}\frac{l}{w}, \quad (3)$$

where $R_{ts}$ is the electrical resistance of the microheater, $\rho$ is the electrical resistivity of gold, $l$ and $w$ are the length and width of the microheater (see Fig. 1) and $R_{sh}$ is the sheet resistance of the film, i.e., the resistance of a portion of film with a square shape, of any size, measured between opposite sides of the square. However, Eq. 3 is not the only constraint that needs to be taken into account. First of all, $R_{ts}$ must be chosen to limit both the voltage and the current needed for a complete shift of the optical signal: $R_{ts} = 50\ \Omega$ is a value that fulfills both these requirements, resulting in a voltage of 5 V and a current of 100 mA, when the microheater dissipates an electrical power of 500 mW [12]. Secondly, the sheet resistance $R_{sh}$ must be chosen sufficiently low to guarantee that the parasitic series resistance introduced by the contact pads is negligible with respect to the microheater resistance $R_{ts}$. Considering a resistor with two contact pads that are approximately square, a maximum sheet resistance $R_{sh} = 0.5\ \Omega/\square$ leads to on-chip parasitics of less than 1 $\Omega$ and, according to Eq. 3, to microheaters featuring a minimum aspect ratio $l/w = 100$ to achieve the desired 50 $\Omega$. These values corresponds to a gold film thickness $h = 50$ nm, calculated by considering the electrical resistivity of bulk gold, namely $\rho = 24.4$ n$\Omega$m. However, the resistivity of a thin film can be considerably higher [21] and, thus, the thickness considered in the following discussion will be $h = 100$ nm, whose compatibility with the design constraints will be demonstrated experimentally in Section III.

## B. Microheater Miniaturization

The previous subsection sets the main constraints on the thermal shifter design. However, it is worth noting that we fixed a minimum value for the ratio $l/w$, but we did not provide any information about the specific values for both the length $l$ and the width $w$. A reduction of both would be very advantageous for the FLW-PIC design, not only because more compact devices can pave the way toward the integration of more complex circuits, but also for other manifest advantages. In fact, a miniaturization in terms of $l$ leads to a reduction of the propagation losses of the waveguides, while a miniaturization in terms of $w$ is important to decrease the power dissipation necessary to induce a given phase shift.

While the first point is trivial to understand, the second one deserves a better explanation. To this aim, let us consider the relation between the electrical power $P_{ts}$ dissipated by the thermal shifter and the phase $\phi$ induced on the MZI. Given reasonable hypothesis on the linearity of the system [12]:

$$\phi = \phi_0 + \alpha P_{ts}, \quad (4)$$

where $\phi_0$ is the phase when no power is injected and $\alpha$ is the linear tuning coefficient of the device. If we consider a microheater length matching the extent of the MZI arms [12]:

$$\alpha = \frac{2n_t}{\lambda k} F = \frac{2n_t}{\lambda k}(F_1 - F_2), \quad (5)$$

where $n_t$ and $k$ are the thermo-optic coefficient and the thermal conductivity of the substrate, $\lambda$ is the wavelength of the photon and $F$ is an efficiency factor that depends on the geometry of the MZI. More specifically, $F$ is the difference of two contributions $F_1$ and $F_2$, that represent how much efficiently each arm is heated up. In order to minimize the power dissipation needed for a given phase shift, $\alpha$ must be designed as high as possible, therefore, it is necessary to place the microheater in thermal contact with the first arm, while isolating the second one.

These considerations can be applied to a planar MZI fabricated at a depth $d$ from the chip surface, with a microheater centered on the first arm and having the second arm at a distance $p$ from the first one. Assuming a pure cylindrical geometry [12], the efficiency factors can be calculated as

$$F_1 = -\log d \quad (6)$$

and

$$F_2 = -\log \sqrt{p^2 + d^2}. \quad (7)$$

However, this model considers the microheater as a wire with negligible transversal dimensions. In order to keep into account also the width $w$, the microheater can be modeled as an infinite series of adjacent wires having an infinitesimal width and providing an infinitesimal contribution to $F$. Mathematically speaking, Eq. 6 and 7 can be extended as

$$F_1 = -\frac{1}{w}\int_{-\frac{w}{2}}^{\frac{w}{2}} \log\left(\sqrt{x^2 + d^2}\right) dx = \\ -\frac{1}{2}\log\left[\left(\frac{w}{2}\right)^2 + d^2\right] - \frac{2d}{w}\arctan\left(\frac{w}{2d}\right) + 1 \quad (8)$$

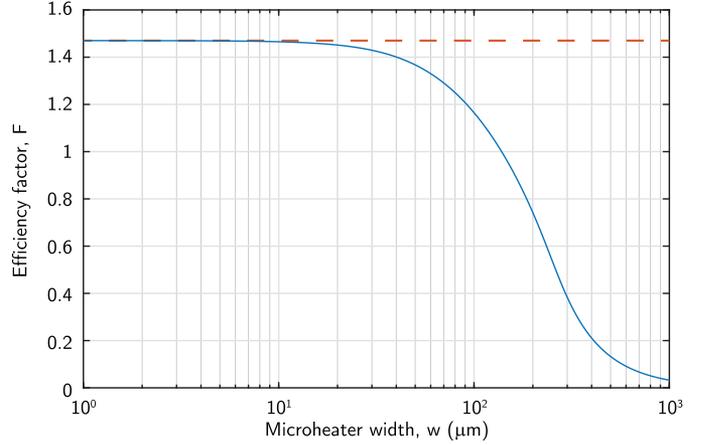

Fig. 2. Efficiency factor $F$ as a function of the microheater width $w$, according to the extended model proposed in this work. A dashed line, with the value produced by the cylindrical model [12], is also reported as reference.

and

$$F_2 = -\frac{1}{w}\int_{-\frac{w}{2}}^{\frac{w}{2}} \log\left[\sqrt{(x+p)^2 + d^2}\right] dx \\
= -\frac{1}{2w}\left(\frac{w}{2} - p\right)\log\left[\left(\frac{w}{2} - p\right)^2 + d^2\right] + \\
-\frac{1}{2w}\left(\frac{w}{2} + p\right)\log\left[\left(\frac{w}{2} + p\right)^2 + d^2\right] + \\
-\frac{d}{w}\left[\arctan\left(\frac{\frac{w}{2} - p}{d}\right) + \arctan\left(\frac{\frac{w}{2} + p}{d}\right)\right] + 1. \quad (9)$$

The efficiency factor $F$ for this extended model is depicted in Fig. 2 as a function of the microheater width $w$. Data are reported for the case of a typical MZI having $d = 30\ \mu$m and $p = 127\ \mu$m. In particular, when $w/2 << p$, the extended model (solid line) produces the same result of the cylindrical one (dashed line) and, thus, the width of the resistor does not affect the power dissipation required for a given phase shift, which is the minimum amount. Conversely, when $w/2$ starts being comparable to $p$, the second arm of the interferometer starts experiencing a more effective heating, that degrades the efficiency factor and, thus, the power dissipation. Eventually, when $w/2 \simeq p$, the resistor heats up both the arms in the same way, producing no shift at all.

In conclusion, in order to minimize the power dissipation needed for a given phase shift, the condition $w/2 << p$ must be always verified. However, it is worth noting that reducing $w/2$ to more than one order of magnitude lower than $p$ makes no sense and, actually, it is detrimental for the overall performance of the FLW-PIC. Indeed, pushing the resistor dimensions beyond this limit results in the same electrical power dissipated on a smaller area and, therefore, in a higher operating temperature, that can impair the stability and the reliability of the microheater.

## III. FABRICATION PROCESS

In order to demonstrate the performance of the new thermal shifters, we purposely developed a new fabrication process, here described along with the preliminary electrical characterization that allowed the optimization of the recipe.

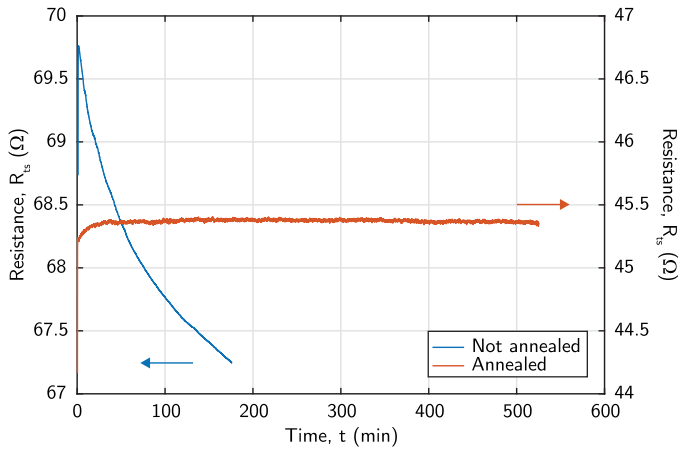

Fig. 3. Resistance $R_{ts}$ plotted as a function of the time $t$ for not annealed and annealed microheaters with same dimensions ($l/w = 120$ and $w = 45$ $\mu$m), when operated at 500 mW.

The deposition of both chromium and gold was performed with a magnetron sputtering system (Leybold LH Z400) on a glass substrate (Corning EAGLE XG alumino-borosilicate, 1.1 mm thick). In light of the previous considerations, a thickness of 2 nm for the chromium layer and a thickness of 100 nm for the gold layer were chosen. After that, the microheaters were patterned ($l/w = 120$ and $w = 45$ $\mu$m) by ablating the film with a Yb:KYW cavity-dumped mode-locked laser (1030 nm wavelength, 300 fs pulse duration, 1 MHz repetition rate) tuned in order to produce pulses of 200 nJ. The light was focused by a 10× objective ($NA = 0.25$) on the chip surface, while the substrate was translated at a speed of 2 mm s$^{-1}$ by a three-axes translation stage (Aerotech ABL10100). To safely isolate the metal features, five successive scans were performed, resulting in ablation lines of 5 $\mu$m width.

Different samples were fabricated to verify reproducibility on a statistically meaningful set of data that includes also depositions performed at different times. A pin header was mounted on each chip, with the pins connected to the contact pads by means of an epoxy electrical conductive glue. After that, each chip was mounted on a metal plate whose temperature was externally controlled and set to 27 °C. The electrical characterization was performed with a Keythley 2612 source meter connected to the resistors through the pin header.

Firstly, we characterized the IV relation of the resistors at low voltage (up to 1 V) in order to prevent any thermal effect on the measurement. A fully ohmic behavior was observed, that allowed us to deduce a sheet resistance $R_{sh} = 0.6$ $\Omega/\square$, more than a factor of 2 higher than the one expected for a bulk material, but consistent with data reported for gold sputtered on a cold glass substrate [21]. Secondly, stability measurements were performed by monitoring the resistance of the microheaters during the operation at 500 mW. As showed in Fig. 3 (blue line), over a time span of only 3 h, the microheater showed an irreversible resistance drop of 2.5 $\Omega$, corresponding to a variation as high as 3.6 %. This behavior can not be considered acceptable for the applications, since it would require the recalibration of the circuit every few hours.

A possible explanation for this issue can be found in the polycrystalline nature of the sputtered gold film: indeed, when a polycrystalline film is heated up, the grains start growing toward the single-crystal state, that is energetically more favorable. Such a modification of the film morphology affects the value of the resistivity, that starts decreasing [21]. For this reason, we introduced a thermal annealing step at the end of the fabrication process, with the aim of having all the grain growth induced by the high thermal budget we provide. The annealing process consists of a rising ramp of 10 °C min$^{-1}$ up to 400 °C, 30 min at this temperature and, in the end, the thermal recovery that lasts no more than 15 h. It is worth pointing out that, being the annealing temperature far from the strain point of the glass substrate (669 °C) [22], this process has no effect on the optical properties of the waveguides, as verified after the integration of the thermal shifters in the complete fabrication process (Section IV).

The characterization was repeated on new samples that underwent the thermal annealing step. Firstly, the low-voltage IV characterization allowed us to measure a sheet resistance $R_{sh} = 0.35$ $\Omega/\square$, almost a factor of 2 lower than before and this time consistent with data reported for sputtering on a hot substrate [21]. Secondly, as reported in Fig. 3 (red line), the new thermal shifters were characterized by a resistance fluctuation lower than 0.1%, calculated on more than 8 h of operation (neglecting the reversible initial thermal transient). This measurement demonstrates that the annealed microheaters require only a single calibration, that can be performed at the beginning of the experiment and that remains valid for at least many hours of operation.

Finally, the thermal annealing should have also a beneficial effect in terms of reliability. Indeed, although we did not investigate this feature in our devices, it is well known from the literature that the grain growth dramatically reduces the electromigration of metal ions [23], that is the main phenomenon leading to the formation of interruptions and voids when a metal line operates at high temperature.

## IV. EXPERIMENTAL ANALYSIS

The new thermal shifters were integrated in an existent FLW process for the fabrication of single-mode waveguides [24]. Test chips were specifically designed and fabricated to characterize the optical performance of reconfigurable MZIs having the geometry considered in Section II. For the optical characterization, coherent light from a laser diode (Thorlabs L785P090, 785 nm wavelength) was coupled inside the waveguides using a 10× objective ($NA = 0.25$). The light was collected at the output using a second objective and its intensity was measured on both the bar and cross outputs of each interferometer by means of two optical power meters (Ophir PD300R-UV). The thermal shifters were biased by connecting them to a bench DC power supply (Keithley 2231A) through the pin header that each chip features.

### A. Transmission Coefficient

A test chip was specifically designed and fabricated to characterize the optical performance of some reconfigurable MZIs featuring thermal shifters with aspect ratio $l/w = 150$ and width $w$ ranging from 16 $\mu$m to 93 $\mu$m. Consistently with the previous discussion, the MZIs were fabricated at a depth $d = 30$ $\mu$m and with an inter-waveguide pitch $p = 127$ $\mu$m, while the thermal shifter fabrication was carried out by depositing, ablating and annealing a thin film of chromium-gold ($R_{sh} = 0.35$ $\Omega/\square$). In particular, the ablation step was not only exploited to electrically isolate the resistors, but also to remove the film between the arms of each MZI, in order to minimize the thermal diffusion and, thus, to maximize the tuning coefficient $\alpha$ attainable with this technology.

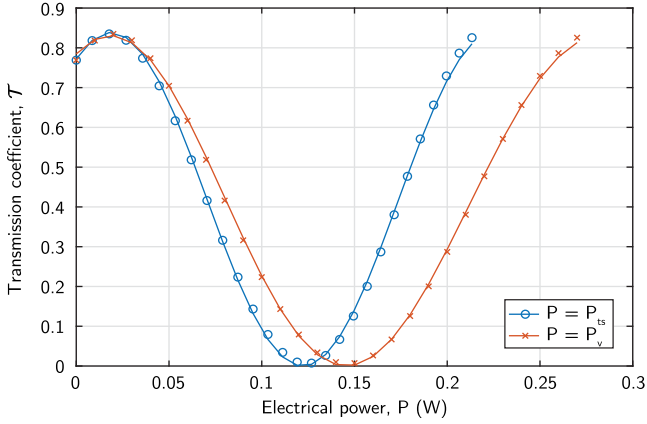

Fig. 4. Transmission coefficient $\mathcal{T}$ measured for the MZI with the most compact thermal shifter ($w = 16$ $\mu$m). Data are plotted as a function of both $P_{ts}$ and $P_v$. Best fitting curves are reported for both the datasets.

The optical response of the fabricated MZIs was characterized as a function of the power dissipated by the corresponding microheater. In particular, Fig. 4 shows the transmission coefficient $\mathcal{T}$ measured at the output of the interferometer having the most compact thermal shifter ($w = 16$ $\mu$m). The same data are plotted as a function of two different variables. In the first dataset (blu circles), $\mathcal{T}$ is reported as a function of the electrical power $P_{ts}$, namely

$$P_{ts} = VI, \quad (10)$$

where $V$ and $I$ are the voltage and the current measured at the external terminals. Nevertheless, controlling the phase shift by acting directly on $P_{ts}$ is not easy. Hence, it is often more convenient to use $V$ as the control variable, redefining the electrical power as

$$P_v = \frac{V^2}{R_0}, \quad (11)$$

where $R_0$ is the resistance of the microheater when the device is not operated. Generally speaking, the resistivity of the thermal shifter has a temperature dependence that can not be neglected and this makes $P_v$ different from $P_{ts}$. In particular, assuming a linear dependence between the microheater resistance $R_{ts}$ and the injected power $P_{ts}$:

$$P_{ts} = \frac{V^2}{R_{ts}} = \frac{V^2}{R_0 + \gamma P_{ts}} \simeq P_v - \frac{\gamma}{R_0} P_v^2, \quad (12)$$

where $\gamma$ is the proportionality coefficient of the relation between resistance and power. As a matter of fact, $P_v$ represents an overestimation of $P_{ts}$, as appreciable also in Fig. 4 by comparing the first dataset with the second one (red crosses), that consists of the same data plotted as a function of $P_v$.

Fig. 4 reports also the corresponding best sinusoidal fits (solid lines). On the one hand, the induced phase $\phi$ exhibits a linear dependence on the electrical power $P_{ts}$ (as predicted by Eq. 4), while, on the other hand, $\phi$ exhibits a nonlinear dependence on $P_v$ (see Eq. 12). Therefore, in both cases the transmission coefficient $\mathcal{T}$ can be modeled as:

$$\mathcal{T} = \frac{\nu}{2}(1 + \sin\phi) = \frac{\nu}{2}[1 + \sin(\phi_0 + \alpha P + \beta P^2)], \quad (13)$$

where $\nu$ is the fringe visibility, $P$ is the independent variable ($P_{ts}$ or $P_v$) and $\beta$ is a coefficient that keeps in account the nonlinear nature of the phase-power relation. Considering the actual power dissipation $P_{ts}$, the quantities $\nu$, $\phi_0$ and $\alpha$ are treated as fitting parameters, while $\beta$ is considered zero. From the $\alpha$ estimation, it is possible to infer the period of the oscillation $P_{2\pi}$, namely the electrical power necessary to induce a $2\pi$ phase shift. The result is $P_{2\pi} = 2\pi/\alpha = 204$ mW, an unprecedented value for a thermally reconfigurable MZI fabricated on a glass substrate. This performance is comparable even to the one obtained from devices having a larger pitch ($p = 254$ $\mu$m) and fabricated with a more complex process based on isolation trenches [18].

Fig. 4 demonstrates also that the power dissipation $P_v$ can be used as a control variable for an accurate operation of the thermal shifter. Indeed, the excellent agreement between model and experimental points confirms that a calibration curve assuming a quadratic dependence (i.e., $\beta \neq 0$) between the phase $\phi$ and the dissipated power $P_v$ is an accurate model for the thermal shifter operation. This allows an easy calibration of the FLW-PIC for the exploitation in practical experiments. In addition, since the dependence on temperature of the resistance is responsible for the different behavior reported in Fig. 4, by analyzing the difference between the two curves we can estimate the maximum operating temperature $T_{2\pi}$ of the microheater. Indeed, the resistance variation is determined by the increment of the transmission coefficient period $P_{2\pi}$: a change from 204 mW to 263 mW, corresponding to a variation of 29 %. Considering a temperature coefficient of resistance (TCR) for the gold of 0.37% °C$^{-1}$, we can deduce a variation of only 78 °C. This limited temperature increase is important not only to prevent any detrimental effect on the reliability, but also to hinder any temperature-induced deformation of the chip, thus maintaining the waveguide coupling stability.

Fig. 4 reports the result of the best fitting only for the microheater having minimal width. However, the estimation of the linear tuning coefficients $\alpha$ was repeated for all the fabricated devices featuring different microheater widths $w$. In particular, the tuning coefficients $\alpha$ can be exploited to validate the theoretical model presented in Section II (Eq. 5, 8 and 9). Since no value is reported in [22] for the thermo-optic coefficient $n_t$ of the substrate, a further best fit was performed on these values, that produced the estimation $n_t = 8.63 \times 10^{-6}$K$^{-1}$. This value is consistent with the ones already employed in the past for very similar substrates [12], [18] and, more generally, with the thermo-optic coefficients reported in the literature for silicate glasses [17].

### B. Thermal and Electrical Crosstalk

Let us now consider a more complex FLW-PIC, consisting of $N$ interferometers integrated in the same substrate. Generally speaking, the operation of a given thermal shifter does not affect only the transmission coefficient of the corresponding MZI, but it can introduce also an undesired phase shift on the other interferometers. The physical origin of this crosstalk can be manifold. As an example, in circuits featuring a very small inter-waveguide pitch, the glass can not be considered a perfect thermal insulator and, as a result, the power injected by a thermal shifter can reach also a nearby device. This phenomenon, usually referred to as thermal crosstalk [15], [18], does not affect the linearity of the phase-power relation, that can be generalized as follows [12]:

$$\phi_i = \phi_{0,i} + \sum_{j=1}^{N} \alpha_{i,j} P_{ts,j}. \quad (14)$$

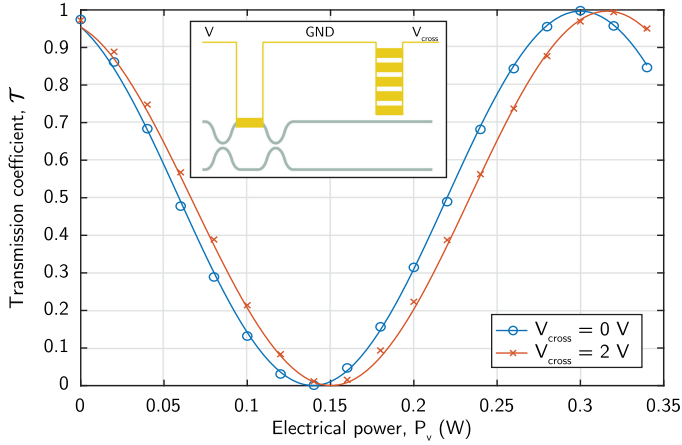

Fig. 5. Transmission coefficient $\mathcal{T}$ as a function of $P_v$, with and without the effect of $V_{cross}$. Best fitting curves are reported for both the datasets. Inset: schematic representation of the circuit designed for the experiment.

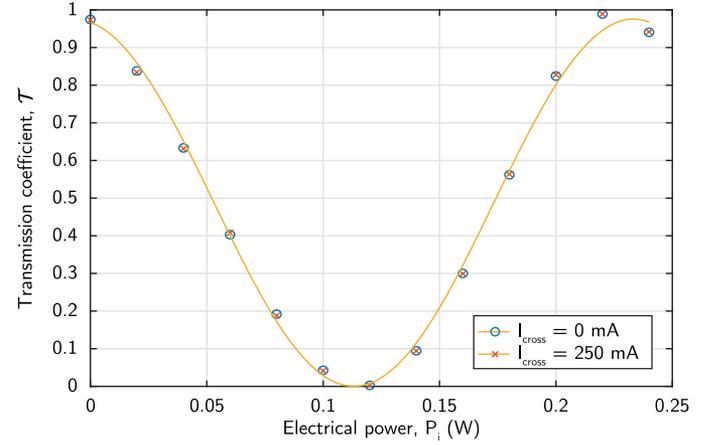

Fig. 6. Transmission coefficient $\mathcal{T}$ as a function of $P_c$, with and without the effect of $I_{cross}$. Both the datasets share the same fitting curve.

However, crosstalk can be introduced also by electrical nonidealities. As an example, let us consider a configuration in which all the resistors share the same ground terminal in order to simplify the metal interconnection layout and to reduce the number of contact pads. The current flowing in the shared terminal is given by the sum of all the bias currents and, if we consider a shared parasitic resistance $R_g$ between the resistors and the actual ground, the voltage drop $V_{ts,j}$ across the j-th microheater can be calculated as

$$V_{ts,j} = V_j - R_g \sum_{k=1}^{N} I_k \simeq V_j - \frac{R_g}{R_{ts}} \sum_{k=1}^{N} V_k. \quad (15)$$

Assuming a voltage-controlled configuration, this effect results in a nonlinear crosstalk among the thermal shifters, that further increases the overestimation of $P_{ts}$. Indeed, neglecting now the temperature dependence of $R_{ts}$, Eq. 15 can be exploited to get

$$P_{ts,j} = \frac{V_{ts,j}^2}{R_{ts}} \simeq P_{v,j} - \frac{2R_g}{R_0} \sum_{k=1}^{N} \sqrt{P_{v,k} P_{v,j}}. \quad (16)$$

In order to study this effect, we fabricated a second test chip, consisting of one reconfigurable MZI and a displaced set of five additional microheaters (see the inset in Fig. 5). We designed all the resistors ($l/w = 125$ and $w = 40$ $\mu$m) with a shared ground terminal. Since the thermal crosstalk between the MZI and the five microheaters is negligible due to the large separation, we observe only the effect of the electrical crosstalk. To this aim,

Fig. 5 reports the transmission coefficient $\mathcal{T}$ as a function of the electrical power $P_v$. The first dataset (blue circles) represents the transmission of the MZI when no other resistor is biased, while, in the second case (red crosses), $\mathcal{T}$ is measured when all the other microheaters are biased with a voltage $V_{cross} = 2$ V, corresponding to a total current $I_{cross} = 250$ mA, which is less than what would be needed to achieve a $2\pi$ phase shift in five MZIs.

In order to demonstrate the consistency between the experimental characterization and the theoretical dissertation, Fig. 5 reports also the corresponding best sinusoidal fits (solid lines). In the first case, the fitting curve follows the model described by Eq. 13, that can be exploited to estimate the parameters $\nu$, $\phi_0$ and $\alpha$, assuming $\beta = 0$ for the sake of simplicity. The second dataset can be fitted by a very similar model, obtained by incorporating Eq. 16 in Eq. 13:

$$\mathcal{T} = \frac{\nu}{2}\left[1 + \sin\left(\phi_0 + \alpha P_v - 2\alpha \frac{R_g}{R_0}\sum_{k=1}^{N}\sqrt{P_{v,k}P_{v,j}}\right)\right], \quad (17)$$

where the only fitting parameter is $R_g$. The result is a parasitic ground resistance $R_g = 0.27$ $\Omega$, a value consistent with the dimensions of the shared metal interconnections.

In conclusion, the experimental analysis shows that a careful design of the metal interconnections is not enough to completely prevent the electrical crosstalk. In order to fully remove this non-linearity, the solution is moving to a current-controlled regime by, firstly, redefining the electrical power as

$$P_c = R_0 I^2 \quad (18)$$

and, secondly, modifying the DC power supply operation, that must be used in constant current mode (i.e., setting the current $I$ instead of the voltage $V$). The same measurements, repeated with the new configuration, are reported in Fig. 6. The experimental points belonging to the two datasets perfectly overlap and, as a consequence, share the same best sinusoidal fit, based on the model of Eq. 13. This result fully demonstrates the effectiveness of driving the thermal shifters in current mode for preventing the nonlinear electrical crosstalk.

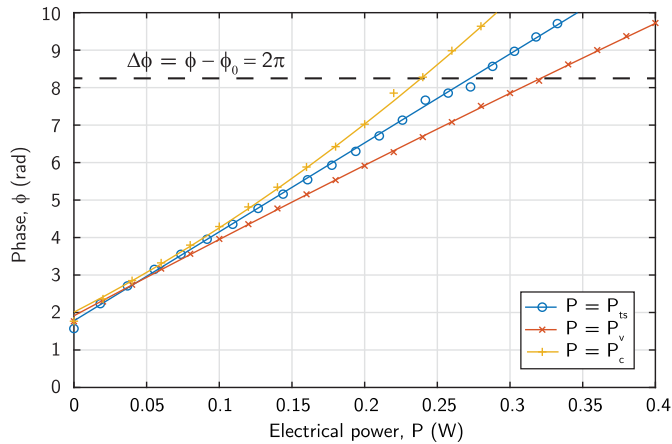

Fig. 7. Induced phase $\phi$ as a function of $P_{ts}$, $P_v$ and $P_c$. A dashed line, indicating the $2\pi$ phase shift, is also reported to highlight the increase or the reduction of the oscillation period when employing $P_v$ or $P_c$.

Finally, it is worth noting that, conversely to what happens with $P_v$, adopting the electrical power $P_c$ leads to an underestimation of the actual power $P_{ts}$, due to the temperature dependence of the resistance $R_{ts}$. Indeed:

$$P_{ts} = R_{ts}I^2 = (R_0 + \gamma P_{ts})I^2 \simeq P_c + \frac{\gamma}{R_0}P_c^2. \qquad (19)$$

This effect is responsible for the contraction of the oscillation period reported in Fig. 6. The different result obtained by employing $P_v$ or $P_c$ is summarized by Fig. 7, that shows the relation between the phase $\phi$ and the different electrical powers for the same device of Figs. 5 and 6. The experimental points for $P = P_v$ and $P = P_c$ are reported along with their best fitting curves, assuming $\beta \neq 0$. By comparing them with the experimental points plotted for $P = P_{ts}$ and fitted assuming $\beta = 0$, it is clear that each driving choice requires a dedicated calibration for accurate control of the thermal shifters.

## V. Conclusion

Thermal phase shifting represents a very simple way to enable the dynamic reconfiguration of a FLW-PIC, but, on the other hand, its implementation in a FLW process is not straightforward. With this work, we have presented a thorough study on the design and fabrication of thermal shifters for FLW-PICs. Thanks to the design rules we have presented, we have been able to fabricate compact microheaters, without compromising neither on the stability nor on the control of the circuit. Furthermore, to the best of our knowledge, we have been able to demonstrate the lowest power dissipation for a $2\pi$ phase shift in a MZI integrated in a glass substrate. Such a performance will allow increasing the number of reconfigurable devices that can be integrated in the same chip before more complex fabrication processes or active cooling devices are needed. In the end, we have also showed that both the crosstalk and the linearity of the FLW-PIC can be improved by replacing the voltage control with a current control of the thermal shifters. The strategies proposed in this paper will be of paramount importance to guarantee an easy calibration and an effective use of the FLW-PIC in practical experiments.

## Acknowledgment

This work was partially performed at PoliFAB, the micro- and nanofabrication facility of Politecnico di Milano (www.polifab.polimi.it). The authors would like to thank the PoliFAB staff (Claudio Somaschini, Alessia Romeo, Marco Asa and Lorenzo Livietti) for the valuable technical support.